%% file: main.tex
\pdfoutput=1
\documentclass[sigconf,screen]{acmart}

\usepackage[T1]{fontenc}
\usepackage[utf8]{inputenc}

\copyrightyear{2021}
\acmYear{2021}
\setcopyright{rightsretained}
\acmConference[IUI '21]{26th International Conference on Intelligent User Interfaces}{April 14--17, 2021}{College Station, TX, USA}
\acmBooktitle{26th International Conference on Intelligent User Interfaces (IUI '21), April 14--17, 2021, College Station, TX, USA}
\acmDOI{10.1145/3397481.3450692}
\acmISBN{978-1-4503-8017-1/21/04}

\usepackage{enumitem} 

\setlist{nolistsep} 
\usepackage{caption}
\usepackage{subcaption}
\usepackage{pifont}
\newcommand{\cmark}{\ding{51}}
\newcommand{\xmark}{\ding{55}}

\newcommand{\tqdb}{\textquotedbl} 

\begin{document}

\title[A Large, Crowdsourced Evaluation of Gesture Generation Systems on Common Data: The GENEA Challenge 2020]{A Large, Crowdsourced Evaluation of Gesture Generation Systems on Common Data: The GENEA Challenge 2020}


\author{Taras Kucherenko}
\email{tarask@kth.se}
\authornote{Equal contribution and joint first authors.}
\affiliation{%
  \institution{Division of Robotics, Perception and Learning, KTH Royal Institute of Technology}
  \city{Stockholm}
  \country{Sweden}
}

\author{Patrik Jonell}
\email{pjjonell@kth.se}
\authornotemark[1]
\affiliation{%
  \institution{Division of Speech, Music and Hearing, KTH Royal Institute of Technology}
  \city{Stockholm}
  \country{Sweden}
}

\author{Youngwoo Yoon}
\authornotemark[1]
\email{youngwoo@etri.re.kr}
\affiliation{%
  \institution{ETRI \& KAIST}
  \city{Daejeon}
  \country{Republic of Korea}
}

\author{Pieter Wolfert}
\email{pieter.wolfert@ugent.be}
\affiliation{%
 \institution{IDLab, Ghent University – imec}
 \city{Ghent}
 \country{Belgium}}

\author{Gustav Eje Henter}
\email{ghe@kth.se}
\affiliation{%
  \institution{Division of Speech, Music and Hearing, KTH Royal Institute of Technology}
  \city{Stockholm}
  \country{Sweden}
}


\begin{abstract}
Co-speech gestures, gestures that accompany speech, play an important role in human communication. Automatic co-speech gesture generation is thus a key enabling technology for embodied conversational agents (ECAs), since humans expect ECAs to be capable of multi-modal communication. Research into gesture generation is rapidly gravitating towards data-driven methods. Unfortunately, individual research efforts in the field are difficult to compare: there are no established benchmarks, and each study tends to use its own dataset, motion visualisation, and evaluation methodology. To address this situation, we launched the GENEA Challenge, a gesture-generation challenge wherein participating teams built automatic gesture-generation systems on a common dataset, and the resulting systems were evaluated in parallel in a large, crowdsourced user study using the same motion-rendering pipeline. Since differences in evaluation outcomes between systems now are solely attributable to differences between the motion-generation methods, this enables benchmarking recent approaches against one another in order to get a better impression of the state of the art in the field. This paper reports on the purpose, design, results, and implications of our challenge.
%
%
\end{abstract}



\settopmatter{printfolios=true} 
\maketitle

\section{Introduction}
This paper is concerned with systems for automatic generation of nonverbal behaviour, and how these can be compared in a fair and systematic way in order to advance the state-of-the-art.
This is of importance as nonverbal behaviour plays a key role in conveying a message in human communication \cite{mcneill1992hand}.
A large part of nonverbal behaviour consists of so called co-speech gestures, spontaneous hand gestures that relate closely to the content of the speech, and that have been shown to improve understanding \cite{holler2018processing}.
Embodied conversational agents (ECAs) benefit from gesticulation, as gesticulation, e.g., improves interaction with social robots \cite{salem2011friendly} and willingness to cooperate with an ECA \cite{salem2013err}. 
Knowledge of how and when to gesture is also needed. This can for example be learned from interaction data; see, e.g., \cite{jonell2020let} and references therein.

Synthetic gestures used to be based on rule-based systems, e.g., \cite{cassell2001beat,salvi2009synface}; see \cite{wagner2014gesture} for a review. These are gradually being supplanted by data-driven approaches, e.g., \cite{bergmann2009GNetIc,levine2010gesture,chiu2015predicting,kucherenko2018data}, with recent work \cite{yoon2019robots,kucherenko2020gesticulator,alexanderson2020style} showing improvements in gesticulation production for ECAs.
However, the results in prior studies on gesture-generation are not directly comparable.
First, prior studies make use of a variety of different evaluation metrics. 
Second, prior studies rely on different data sources, and train their models on these different sources. 
Lastly, visualisations of their generated gestures have different avatars and production values, which can obscure the quality of the underlying gesture-generation approach.
All these differences are, however, external to the actual methods that drive the gesture generation. 

In this paper, we present the GENEA Challenge 2020,\footnote{GENEA stands for ``Generation and Evaluation of Non-verbal Behaviour for Embodied Agents''. The paper extends a preliminary report, \cite{kucherenko2020genea} (not peer reviewed), presented at the GENEA Workshop associated with the challenge.} the first joint gesture-generation challenge that controls for all previous sources of between-paper variation, by providing a common dataset for building gesture-generation systems, along with common evaluation standards and a shared visualisation procedure.
The aim of the challenge is not to select the best team -- it is not a contest, nor a competition -- but to be able to compare different approaches and outcomes.
This makes it possible to assess and advance the state of the art in gesture generation, and measure the gap between it and natural co-speech gestures.
Comparing the different methods and their performance also helps identify what matters most in gesture generation, and where the bottlenecks are.
Challenge participants benefit by working on the same problem together with researchers interested in the same topic, strengthening the research community, and get an opportunity to compare their systems to other competitive systems in a large and carefully-executed joint evaluation. 




Our concrete contributions are:
\begin{enumerate}
    \item Jointly evaluating several state-of-the-art gesture-generation models on a common dataset using a common 3D model and rendering method.
    \item Two large-scale user studies assessing human-likeness and appropriateness of submitted motion.
    \item Providing open code and and high-quality data -- comprising the pre-processed, multimodal training and test datasets, the standardised visualisation, a large number of subjective responses, and evaluation and analysis using open standards and code -- in the spirit of reproducible research.
    \item Bringing researchers together in order to advance the state-of-the-art in gesture generation, and enabling future research to compare and benchmark against systems from the challenge.
\end{enumerate}

The remainder of this paper first presents prior work in terms of gesture-evaluation practices (and their shortcomings) and discusses how challenges have helped in other fields. We then describe the challenge setup and its results, and finally turn to consider the implications for future challenges and gesture generation as a whole. 


\section{Related work}
Most previous work proposing new gesture-generation methods incorporates an evaluation to support the merits of their method. Human gesture perception is highly subjective, and there are currently no widely accepted objective measures of gesture perception,
so most publications have conducted human assessments instead.
However, previous subjective evaluations, as reviewed in \cite{wolfert2021review}, have several drawbacks, with major ones being the coverage of systems being compared and the scale of the studies.
Like in \cite{sadoughi2019speech,kucherenko2020gesticulator,kucherenko2021moving,alexanderson2020style}, proposed models are at most compared to one or two prior approaches (often a highly similar baseline) or possibly only to ablated versions of the same model.
A large number of studies do not compare their outcomes with other methods at all.
This creates an insular landscape where particular model families only are applied to particular datasets, and never contrasted against one another.
As for scale, large evaluations are expensive, and studies may not be be able to recruit enough participants, thus leaving the differences between many pairs of studied systems unresolved and not statistically significant (cf.\ \cite{yoon2019robots,yoon2020speech}).
Questionnaires, which are one popular evaluation methodology (cf.\ \cite{salem2012generation,ishi2018speech,bergmann2010individualized}) 
demand a lot of time and cognitive effort even before scaling up.
In addition, the items used in questionnaires differs across studies and the set of questions used is often not standardised.

Sometimes, evaluations fail to anchor system performance against natural (``ground truth'') motion from their database, e.g., \cite{salem2012generation,ishii2018generating,le2012evaluating}.
Another significant difference between studies is how generated motion is visualised, where some prior work
(e.g., \cite{wolfert2019should,kucherenko2019analyzing}) displays motion through stick figures, or applies it to a physical agent (e.g., \cite{salem2012generation,ishi2018speech}).
Neither of these may allow the same expressiveness or range of motion as 3D-rendered avatars in, e.g., \cite{alexanderson2020style,kucherenko2020gesticulator}.



Although there is no directly related work on challenges that benchmark co-speech gestures in ECAs, other fields have done well using challenges to standardise evaluation techniques, establish benchmarks, and track and evolve the state of the art. 
For example, the Blizzard Challenges have since their inception in 2005 (see \cite{black2005blizzard}) helped advance text-to-speech (TTS) technology and identified subtle but robust trends in the specific strengths and weaknesses in different speech-synthesis paradigms \cite{king2014measuring}. 
These challenges are open to both academia and industry.
Participants are provided a common dataset of speech audio and associated text transcriptions, and use these to build a synthetic voice. 
The resulting voices are then evaluated in a large, joint evaluation. Challenge data, evaluation stimuli, and subjective ratings remain available after the challenge, and have been widely used both for benchmarking subsequent TTS systems, e.g., \cite{szekely2012evaluating,charfuelan2013expressive}, and for doing research on the perception of natural and artificial speech, e.g., \cite{moller2010comparison,yoshimura2016hierarchical,mittag2020deep,saratxaga2016synthetic,govender2019using}.

Challenges are also actively used in the computer-vision community, for instance for benchmarking purposes. 
Recent CLIC\ \cite{clic2020} and NTIRE\ \cite{ntire2020} challenges, for example, compared systems for image compression and super-resolution respectively, also incorporating subjective human assessments similar to the challenge described in this paper (although they used a MOS-like setup, which has been found to be less efficient than the side-by-side evaluation methodology we employ \cite{ribeiro2015perceptual}).
This addresses the over-reliance on objective metrics in computer-vision evaluation, which, just like in speech quality and gesture generation, do not always align with human perception.
Inspired by the successes of challenges in other field of study, we conducted the first challenge in the field of gesture generation.


\section{Task}
Our challenge focussed on data-driven gesture generation. We pose the problem of speech-driven gesture generation as follows: given input speech features $\boldsymbol{s}$ -- which could involve either an audio waveform (a sequence of pressure samples) or text (a word sequence) or the combination of the two -- the task is to generate a corresponding pose sequence $\boldsymbol{\hat{g}}$ describing gesture motion that an agent might perform while uttering this speech.
To enable direct comparison of different data-driven gesture-generation methods, all methods evaluated in the challenge were trained of the same gesture-speech dataset and their motion visualised using the same virtual avatar and rendering pipeline.


\subsection{Dataset}

We based the challenge on the Trinity Gesture Dataset \cite{ferstl2018investigating}, comprising 244 min of audio and motion-capture recordings of a male actor speaking freely on a variety of topics.
This is one of the largest datasets of parallel speech and 3D motion (in joint-angle space) publicly available in the English language.
We removed lower-body data, retaining 15 upper-body joints out of the original 69.
Finger motion was also removed due to poor capture quality.

To obtain verbal information from the speech, we first transcribed the audio recordings using \href{https://cloud.google.com/speech-to-text/}{Google Cloud automatic speech recognition} (ASR),
followed by a thorough manual review to correct recognition errors and add punctuation for both the training and test parts of the dataset. All names of non-fictive persons were removed and replaced by unique tokens in the transcriptions.

Before releasing the data to challenge participants, it was split into training data (3 h and 40 min) and test data (20 min), with only the training data initially being shared with the participants.
Both these data subsets have since been made publicly available in the original dataset repository at  \href{https://trinityspeechgesture.scss.tcd.ie/}{trinityspeechgesture.scss.tcd.ie}.



\subsection{Challenge rules}

Each participating team could only submit one system for evaluation. 
%
%
As for timeline, the speech-motion training data was released to participants on July 1, 2020. Test input speech (but not motion output) was released to participants on August 7, with participants requested to submit their generated gesture motion for the test input speech on or before August 15. 
The joint evaluation took place after the generated gestures were submitted.


Synthetic gesture motion was required to be submitted at 20 frames per second (fps) in a format otherwise identical to that used by the challenge training data.
To prevent optimising for the specific evaluation used in the challenge and to encourage motion generation approaches with long-term stability, participants were asked to synthesise motions for 20 min of test speech in long contiguous segments, from which a subset of clips were extracted for the user studies, similar to many Blizzard Challenges. Manual tweaking of the output motion was not allowed, since the idea was to evaluate how systems would perform in an unattended setting.


\section{Systems and teams}
We recruited challenge participants from a public call for participation. Sixteen teams signed up for the challenge, and we distributed the dataset and baseline implementations to all of them. Five teams completed the challenge and the other teams were not able to submit results for evaluation. Two of the withdrawing teams explained it was (in one case) due to reduced manpower for completing the challenge and (in the other) due to unsatisfactory results. There were no reported withdrawals due to the challenge data or task.

The challenge evaluation contained 9 different \emph{conditions} or \emph{systems}: 2 toplines that represent human-quality gesture motions, 2 previously published \emph{baselines}, and 5 challenge \emph{entries/submissions}. Table \ref{tab:conditions} lists all conditions, together with participating team names and (abbreviated) affiliations. Following the practice established by the Blizzard Challenge, we anonymised the teams in the present paper, by not revealing which team was assigned which ID, but individual teams are free to disclose their ID if they wish.
Papers from each team describing their submitted systems in detail are collected in the proceedings of the GENEA Workshop 2020.\footnote{Available at \href{https://zenodo.org/communities/genea2020/}{zenodo.org/communities/genea2020}.}

The two toplines were:
\begin{itemize}
\item[\textbf{N}] Natural motion capture from the actor for the input speech segment in question. Surpassing this system would essentially entail superhuman performance.
\item[\textbf{M}] \emph{Mismatched} natural motion capture from the actor, corresponding to another speech segment than that played together with the video. This was accomplished by permuting the motion segments from condition N in such a way that no segments remained in its original position. This represents the performance attainable by a system that produces very human-like motion (same as N, so a topline), but whose behaviour is completely unrelated to the speech (and thus can be considered as a bottom line in terms of motion appropriateness for the speech).
\end{itemize}

Since there has been no previous general study that compares systems to each other and what the state of the art is, it is hard to identify the ``best'' baseline systems to use. Therefore the choice was more subjective and based on code availability, with the two baseline systems chosen from recent data-driven gesture-generation papers that had their code available and were easy to reproduce. These were:
\begin{itemize}
\item[\textbf{BA}] The system from \cite{kucherenko2019analyzing}, which only takes speech audio into account when generating system output. This model uses a chain of two neural networks: one maps from speech to pose representation and another decodes representation to pose, generating motion frame by frame by sliding a window over the speech input.
\item[\textbf{BT}] The system from \cite{yoon2019robots}, which only takes text transcript information (which includes word timing information) into account when generating system output. This model consists of an encoder for text understanding and a decoder for frame-by-frame pose generation.
\end{itemize}

The original authors of the baseline systems updated their methods and code to perform well on the challenge material.
In BA, the representation of upper-body poses in the challenge dataset was different from the data used in the original publication and hence a new hyperparameter search was conducted to find optimal hyperparameters. Another change was that the resulting motion was represented using the exponential map \cite{grassia1998practical} and was smoothed using a Savitzky–Golay filter \cite{savitzky1964smoothing} with window length 9 and polynomial order 3. 

In BT, the representation of upper-body poses in the challenge dataset was different to that of the TED dataset used in the original publication. Accordingly, the pose representation was changed from 2D Cartesian coordinates of 8 upper-body joints to 3${\times}$3 rotational matrices for each of 15 joints. The data dimension for a pose was 135 (3${\times}$3${\times}$15). The number of layers and loss function were the same as in the original paper. The hyperparameters of learning rate and loss term weights were adjusted manually. Also, pretrained FastText word vectors \cite{bojanowski2016enriching} were used instead of GloVe \cite{pennington2014glove}. 
\input{conditions_table.tex}

Source code and hyperparameters for both baseline systems are available on GitHub.\footnote{BA: \href{https://github.com/GestureGeneration/Speech_driven_gesture_generation_with_autoencoder/tree/GENEA_2020}{github.com/GestureGeneration/Speech\_driven\_gesture\_generation\_with\_ autoencoder/tree/GENEA\_2020}\\{}\hphantom{\textsuperscript{3}}BT: \href{https://github.com/youngwoo-yoon/Co-Speech_Gesture_Generation}{github.com/youngwoo-yoon/Co-Speech\_Gesture\_Generation}} These implementations and hyperparameters were also made available to participating teams during the challenge.

We also considered including a re-implementation of the system from Ginosar et al.\ \cite{ginosar2019learning} as a third baseline, but this was dropped due to unsatisfactory results.
This might be due to the challenge dataset being smaller than needed for this method, or due to difficulties with tuning the particular implementation we used.

\section{Evaluation}
We conducted a large-scale, crowdsourced, joint evaluation of gesture motion from the 
nine conditions in Table \ref{tab:conditions} in parallel using a within-subject design (i.e., every rater was exposed to and evaluated all conditions).
The systems were evaluated in terms of the human-likeness of the gesture motion itself, as well as the appropriateness of the gestures for a given input speech.
Jonell \& Kucherenko et al.\ \cite{jonell2020iva_crowd} recently found that the results from crowdsourcing evaluations were not significantly different from in-lab evaluations in terms of results and consistency. We therefore adopted an entirely crowdsourced approach, as opposed to for example the Blizzard Challenge, which has used a mixed approach. Attention checks were used to exclude participants that were not paying attention, as detailed in Section\ \ref{ssec:studydesign}.

\subsection{Stimuli}
Prior to motion being submitted, the organisers selected 40 non-overlapping speech segments from the test inputs (average segment duration 10 s) to use in the user-study evaluation. 
These speech segments, which were not revealed to participants, were selected across the test inputs to be full and/or coherent phrases. 
The motion from the corresponding intervals in the BVH files submitted by participating teams was extracted and converted to a motion video clip using the visualisation server provided to participants (see Section\ \ref{subs:vis_ser}), albeit at a higher resolution of 960$\times$540 this time.
\begin{figure}[!t]
\centering
    \includegraphics[width=\columnwidth]{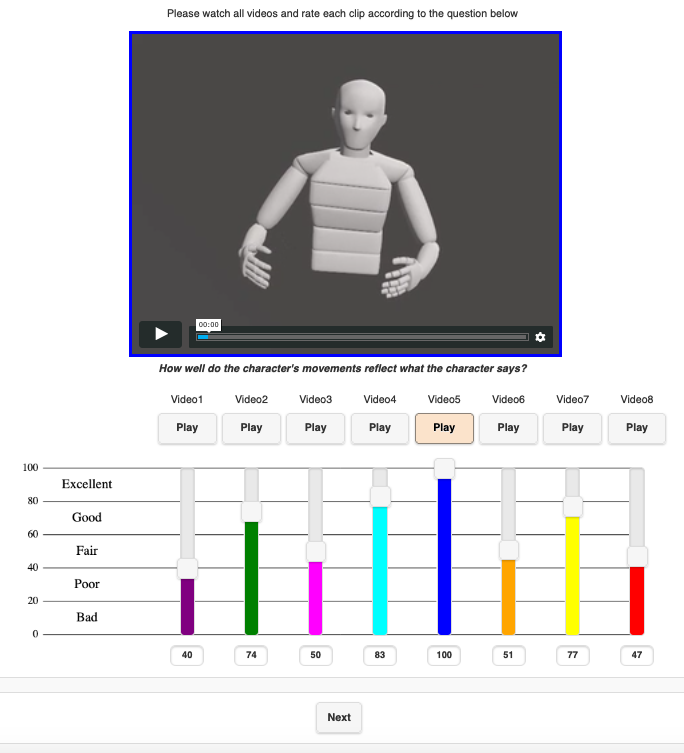}
\caption{Screenshot of the rating interface from the evaluation. The question asked in the image (``How well do the character's movements reflect what the character says?'') originates from \cite{jonell2021hemvip}, and was changed for each of the two evaluations in this paper.}
\label{fig:evaluation_interface}
\end{figure}

We used the same virtual avatar for all renderings during the challenge and the evaluation. The avatar can be seen in Figure\ \ref{fig:evaluation_interface}. The avatar originally had 69 joints (full body including fingers) but only 15 joints, corresponding to the upper body and no fingers, were used for the challenge. Since hand and finger data had been omitted, these body parts were assigned a static pose, in which the hands were lightly cupped (again, see Figure\ \ref{fig:evaluation_interface}).

\label{subs:vis_ser}
We also developed a visualisation server that enabled all participating teams to produce gesture-motion visualisations identical (except in resolution) to the video stimuli evaluated in the challenge. This was implemented using a Python-based web server which interfaced \href{https://www.blender.org}{Blender} 2.83.
Participants would send a 
send a 20 fps BVH file to the visualisation server,
and these files were then processed as quickly as possible into videos visualising the motion on the avatar, in the order they came in.
The same server was also used to render the final stimuli, but with the resolution increased to 960$\times$540 instead of 480$\times$270. (The lower resolution was used during the main part of the challenge to increase performance and throughput of the server, since 16 teams initially took part.)
The visualisation server code is provided at \href{https://github.com/jonepatr/genea_visualizer}{github.com/jonepatr/genea\_visualizer}.

\subsection{Evaluation interface}
In order to efficiently evaluate a large number of relatively similarly-performing systems in parallel, we used a methodology inspired by the MUSHRA (MUltiple Stimuli with Hidden Reference and Anchor) test standard for audio-quality evaluation \cite{itu2015method} from the International Telecommunication Union (ITU). However, there are a number of differences between the MUSHRA standard and our evaluation, e.g., our use of video rather than audio and the omission of a designated reference and a low-end anchor, which correspond to the letters R and A in the original acronym.

Figure\ \ref{fig:evaluation_interface} shows an example of the user interface used for the evaluation. The participants were first met with a screen with instructions and how to use the evaluation interface. They were then presented with 10 pages, where on each page they would compare and evaluate motion stimuli from all toplines, baselines, and most submitted systems, all for/with the same speech. It was possible for participants to return to previous conditions and change their rating after seeing other examples. Lastly they were presented with a page asking for demographics and their experience of the test.
As can be seen in the figure, the 100-point rating scale was anchored by dividing it into successive 20-point intervals labelled (from best to worst) ``Excellent'', ``Good'', ``Fair'', ``Poor'', and ``Bad''. These labels were based on those associated with the 5-point scale used for Mean Opinion Score (MOS) \cite{itu1996telephone} tests, another evaluation standard developed by the ITU.

For a detailed explanation of the evaluation interface we refer the reader to \cite{jonell2021hemvip}, which introduced and validated the evaluation paradigm for gesture-motion stimuli.

\subsection{Study design}
\label{ssec:studydesign}
Each study was balanced
such that each segment appeared on pages 1 through 10 with approximately equal frequency across all raters (segment order), and each condition was associated with each slider with approximately equal frequency across all pages (condition order). For any given participant and study, each page would use different speech segments. Every page would contain condition N and (where relevant) condition M, but one other condition was randomly omitted from each page to limit the maximum number of sliders on a page to 8 or 7, depending on the study.

Three attention checks were incorporated into the pages for each study participant. These either displayed a brief text message over the gesticulating avatar reading ``Attention! Please rate this video XX.'', or they temporarily replaced the audio with a synthetic voice speaking the same message. XX would be a number from 5 to 95, and the participant had to set the corresponding slider to the requested value, plus or minus 3, to pass the attention check. The numbers 13 through 19, as well as multiples of 10 from 30 to 90, were not used for attention checks due to their acoustic ambiguity. Which sliders on which pages that were used for attention check was uniformly random, except that no page had more than one attention check, and condition N and M were never replaced by attention checks.

We evaluated two aspects of the gesture motion, each in a separate study:
\begin{description}
\item[Human-likeness] This study asked participants to rate ``How human-like does the gesture motion appear?'', with the intention of measuring the quality of the generated motion while ignoring its link to the input speech. This study did not include speech in stimulus videos and only used text-based attention checks (all videos were silent).
\item[Appropriateness] This study asked participants to rate ``How appropriate are the gestures for the speech?'' This was intended to investigate the perceived link between motion and speech (both in terms of rhythm/timing and semantics), ignoring motion quality as much as possible. This study contained speech audio in the stimuli, and each participant had to pass one text-based and two audio-based attention checks.
\end{description}

\subsection{Test-participant recruitment}
\label{ssec:recruitment}

Study participants were recruited through the crowdsourcing platform \href{https://www.prolific.co/}{Prolific} (formerly Prolific Academic), restricted to a set of English-speaking countries (UK, IE, USA, CAN, AUS, NZ). There was no requirement to be a native speaker of English, since Prolific does not support screening participants based on that criterion. A participant could take either study or both studies, but not more than once each. Participants were remunerated 5.75 GBP for completing the human-likeness study (median time 33 min) and 6.50 GBP for the appropriateness study (median time 34 min).


\subsection{Objective evaluation metrics}
\label{subs:obj_metrics}

Since subjective evaluation is costly and time-consuming it would be beneficial for the field to agree on meaningful objective evaluations to use. As a step in this direction we consider two numerical measures previously used to evaluate co-speech gestures, namely
average jerk and distance between gesture speed (i.e., absolute velocity) histograms.


\subsubsection{Average jerk}
The third time derivative of the joint positions is called \emph{jerk}.
Average jerk is commonly used to quantify motion smoothness \cite{morasso1981spatial,uno1989formation,kucherenko2019analyzing}.
We report average values of absolute jerk (defined using finite differences) across different motion segments. A perfectly natural system should have average jerk very similar to natural motion.

\subsubsection{Comparing speed histograms}
The distance between speed histograms has also been used to evaluate gesture quality \cite{kucherenko2019analyzing,kucherenko2020gesticulator}, since well-trained models should produce motion with similar properties to that of the actor it was trained on.
In particular, it should have a similar motion-speed profile for any given joint.
To evaluate this similarity we calculate speed-distribution histograms for all systems and compare them to the speed distribution of natural motion (condition N) by computing the Hellinger distance \cite{nikulin2001hellinger}, $H(\boldsymbol{h}^{(1)},\, \boldsymbol{h}^{(2)}) = \sqrt{1 - \sum_{i}{\sqrt{h^{(1)}_i \cdot h^{(2)}_i}}}$, between the histograms $\boldsymbol{h}^{(1)}$ and $\boldsymbol{h}^{(2)}$.
Lower distance is better.

For both of the objective evaluations above the motion was first converted from joint angles to 3D coordinates. The code for the numerical evaluations has been made publicly available to enhance reproducibility.\footnote{See \href{https://github.com/Svito-zar/genea_numerical_evaluations}{github.com/Svito-zar/genea\_numerical\_evaluations}.}


\section{Results and findings of the challenge evaluation}%
\label{sec:results}
This section describes and discusses the results of the subjective and objective evaluations. 
First, Section\ \ref{ssec:demographics} introduces demographic and other information gathered from the recruited participants.
Section\ \ref{ssec:subjectiveresults} then reports the results of the subjective evaluation of challenge conditions, which also are visualised in a number of different figures.
Section\ \ref{ssec:objectiveresults} complements the subjective findings with results on the objective measures introduced in Section\ \ref{subs:obj_metrics}.
Section\ \ref{ssec:discussion_of_results} provides a discussion of the results obtained in the challenge evaluation.

\subsection{Data on test participants}
\label{ssec:demographics}
Each user study recruited 125 participants that passed all attention checks they encountered.
In the human-likeness study, average reported participant age was 31.5 years (standard deviation 10.7), with 66 men, 57 women, and 2 others. 
We asked participants on which continent they lived, and 69 participants were from Europe, 1 from Africa, 48 from North America, 2 from South America, and 5 from Asia. 
In the appropriateness study, average age was 31.1 years (standard deviation 11.7), with 60 men, 64 women, and 1 other.
78 participants reported residing in Europe, 1 in Africa, 39 in North America, 3 in Asia, and 4 in Oceania.
Each study had 116 native and 9 non-native speakers of English.

23 test-takers in the human-likeness study and 40 test-takers in the appropriateness study did not pass all attention checks. These test-takers were not part of the 125 participants analysed. Scores from sliders used for attention checks were also omitted, leaving in total 8,375 and 9,625 ratings that were analysed in each of the two respective studies.
The median successful completion time for the main part of the study
was 24 min for the human-likeness study and 27 min for the appropriateness study, with the shortest successful completion time being 12 min in both studies.
These figures exclude reading instructions and answering the post-test questionnaire, unlike the timings in Section\ \ref{ssec:recruitment}.


\subsection{Analysis and results of subjective evaluation}
\label{ssec:subjectiveresults}
\begin{figure*}[tp!]
\centering
  \begin{subfigure}[b]{\columnwidth}
    \centering\includegraphics[width=\columnwidth]{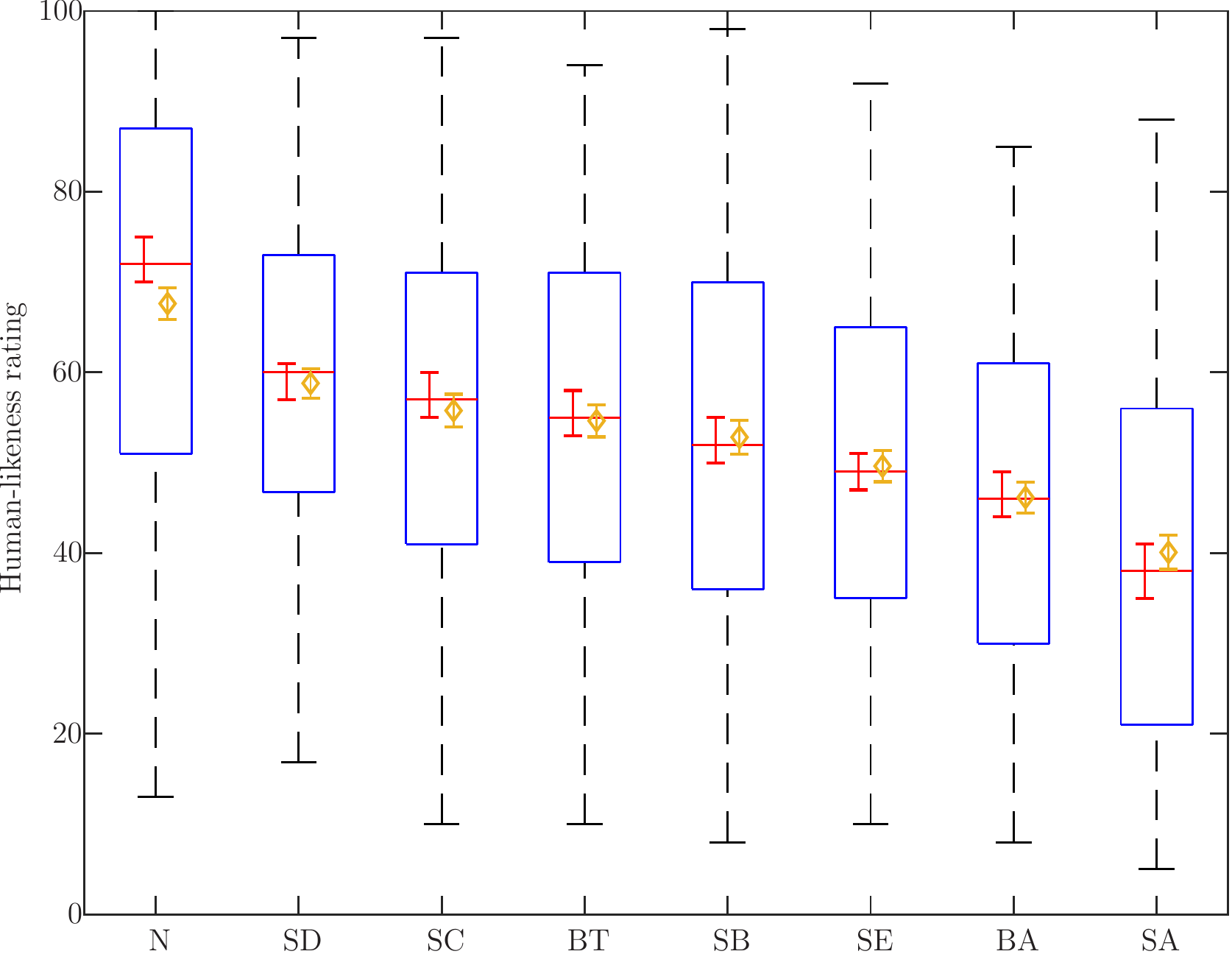}
    \caption{Human-likeness ratings}
    \label{sfig:humanboxplot}
  \end{subfigure}
  \hfill
  \begin{subfigure}[b]{\columnwidth}
    \centering\includegraphics[width=\columnwidth]{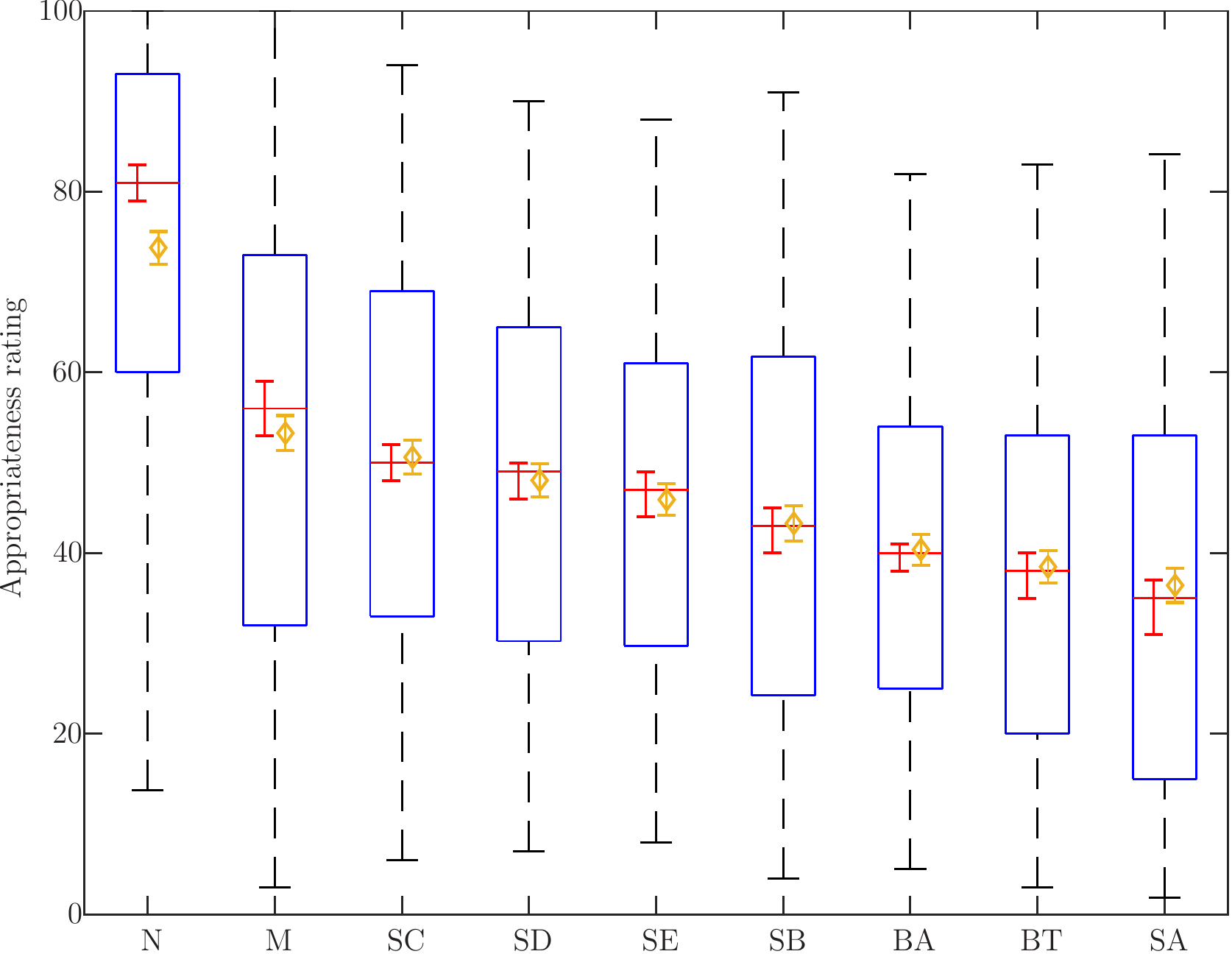}
    \caption{Appropriateness ratings}
    \label{sfig:appboxplot}
  \end{subfigure}
\caption{Box plots visualising the ratings distribution in the two studies. Red bars are the median ratings (each with a 0.01 confidence interval); yellow diamonds are mean ratings (also with a 0.01 confidence interval). Box edges are at 25 and 75 percentiles, while whiskers cover 95\% of all ratings for each system. Conditions are ordered descending by sample median, which leads to a different order in each of the two plots.}
\label{fig:boxplots}
\end{figure*}%
Summary statistics (sample median and sample mean) for all conditions in each of the two studies are shown in Table \ref{tab:results} (see page 8), together with a 99\% confidence interval for the true median/mean.
The confidence intervals were computed either using a Gaussian assumption for the means (i.e., with Student's $t$-distribution cdf, and rounded outward to ensure sufficient coverage), or using order statistics for the median (leverages the binomial distribution cdf, cf. \cite{hahn1991statistical}).


The ratings distributions in the two studies are further visualised through box plots in Figure\ \ref{fig:boxplots}.
The distributions are seen to be quite broad.
This is common in MUSHRA-like evaluations, since the range of numbers not only reflects differences between systems, but also extraneous variation, e.g., between stimuli, in individual preferences, and in how critical different raters are in their judgements.
In contrast, the plotted confidence intervals are seen to be quite narrow, due to the large number of ratings collected for each condition.
\begin{figure*}[tp!]
\centering
  \begin{subfigure}[b]{\columnwidth}
    \centering\includegraphics[width=\columnwidth]{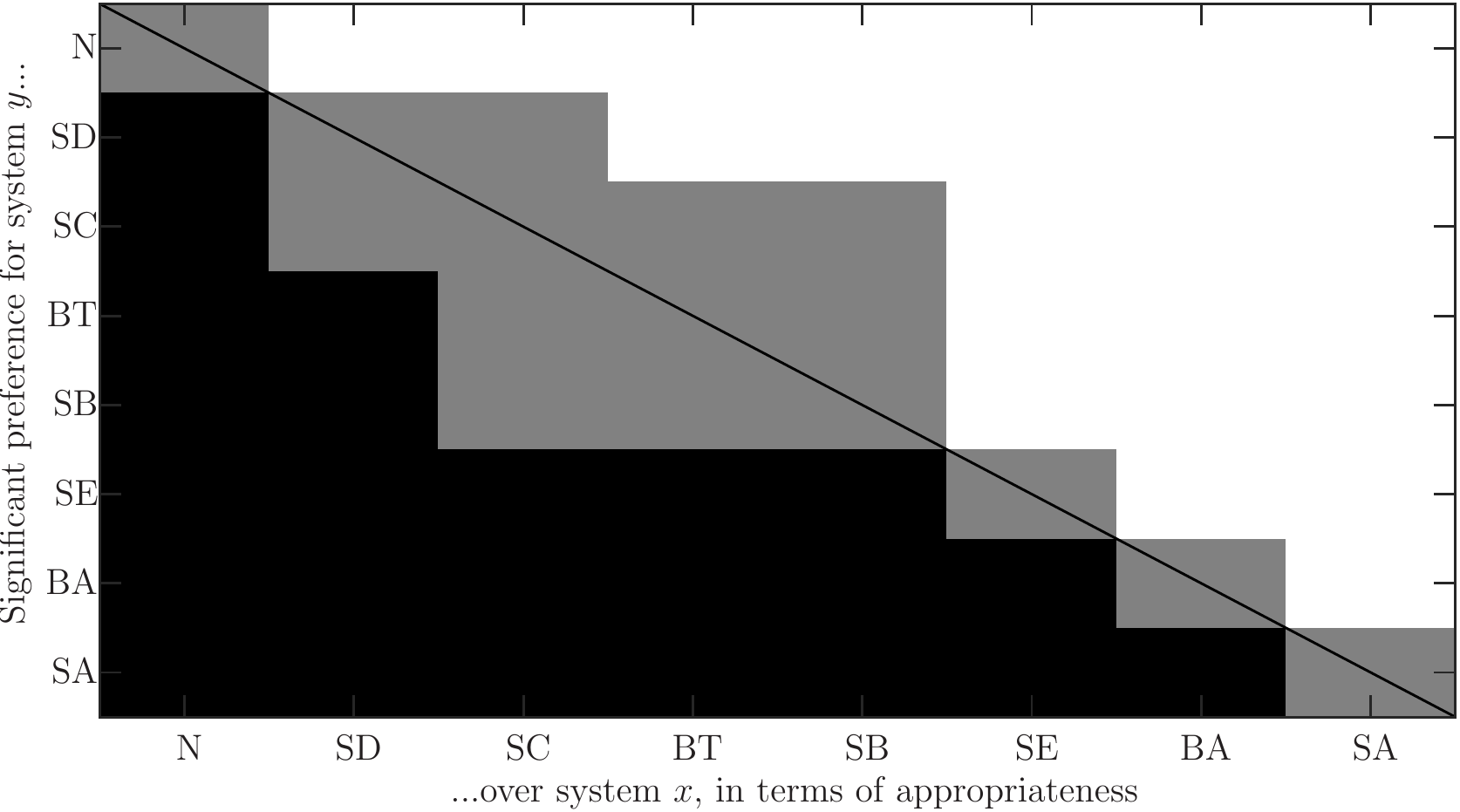}
    \caption{Human-likeness study}
    \label{sfig:humansignificance}
  \end{subfigure}%
  \hfill
  \begin{subfigure}[b]{\columnwidth}
    \centering\includegraphics[width=\columnwidth]{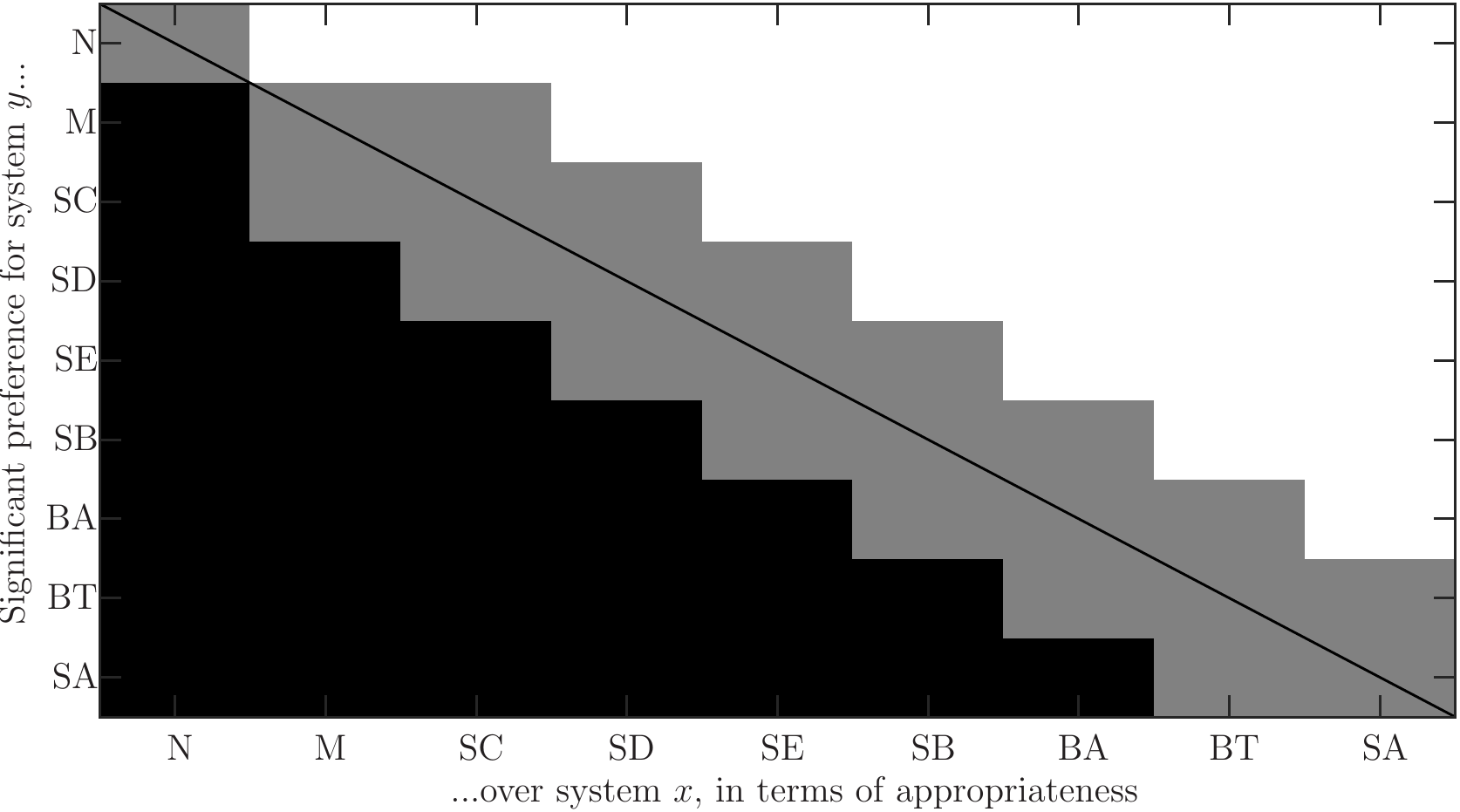}
    \caption{Appropriateness study}
    \label{sfig:appsignificance}
  \end{subfigure}
\caption{Significance of pairwise differences between conditions. White means that the condition listed on the $y$-axis rated significantly above the condition on the $x$-axis, black means the opposite ($y$ rated below $x$), and grey means no statistically significant difference at the 0.01 level after Holm-Bonferroni correction. Conditions are listed in the same order as in Figure\ \ref{fig:boxplots}, which is different for each of the two studies.%
}
\label{fig:significance}
\end{figure*}
\begin{figure*}[t!]
\centering
\includegraphics[width=0.8\textwidth]{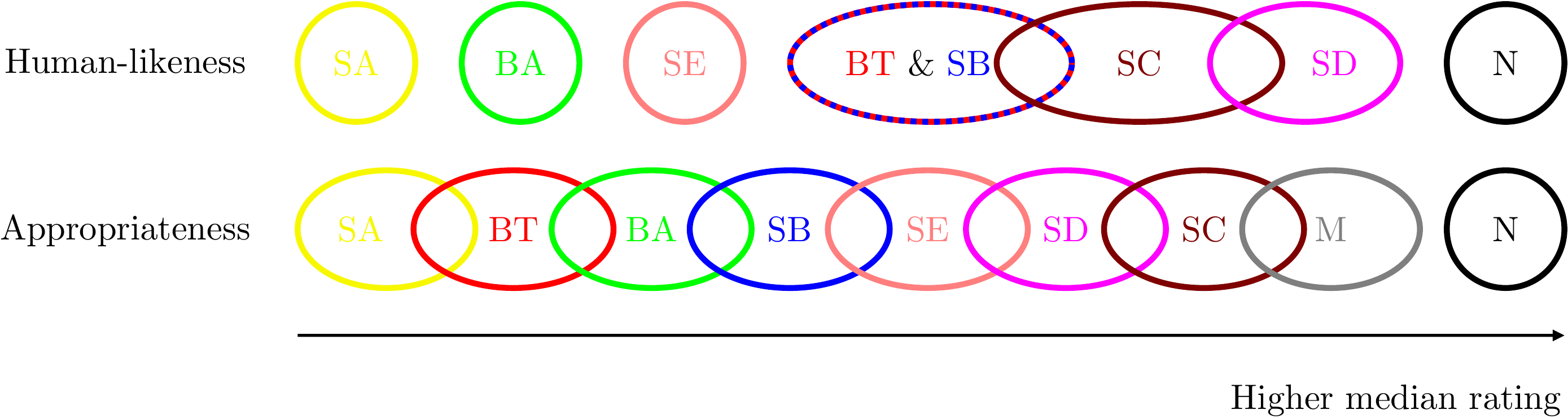}
\caption{Partial ordering between conditions in the two studies. Each condition is an ellipse; overlapping or (in one case) coinciding ellipses signify that the corresponding conditions were not statistically significantly different in the evaluation. The diagram was inspired by \cite{wester2016analysis} with colours adapted from \cite{boynton1989eleven}. There is no scale on the axis since the figure visualises ordinal information only.}
\label{fig:ordering}
\end{figure*}

Despite the wide range of the distributions, the fact that the conditions were rated in parallel on each page enables using pairwise statistical tests to factor out many of the above sources of variation.
To analyse the significance of differences in sample median between different conditions, we applied two-sided pairwise Wilcoxon signed-rank tests to all pairs of distinct conditions in each study.
This closely follows the analysis methodology used throughout recent Blizzard Challenges.
(Unlike Student's $t$-test, this test does not assume that rating differences follow a Gaussian distribution, which would likely be inappropriate, as we can see from the box plots in Figure\ \ref{fig:boxplots} that ratings distributions are skewed and thus non-Gaussian.)
For each condition pair, only pages for which both conditions were assigned valid scores were included in the analysis. (Recall that not all systems were scored on all pages due to the limited number of sliders and the presence of attention checks.)
This meant that every statistical significance test was based on at least 796 pairs of valid ratings in each of the studies.
The $p$-values computed in the significance tests were adjusted for multiple comparisons using the Holm-Bonferroni method \cite{holm1979simple} (which is uniformly more powerful than regular Bonferroni correction)
in each of the two studies.
This statistical analysis found all but 4 out of 28 condition pairs to be significantly different in the human-likeness study, which the corresponding numbers being 7 out of 36 condition pairs in the appropriateness study, all at the level $\alpha=0.01$.
Which conditions that were found to be rated significantly above or below which other conditions in the two studies is visualised in Figure\ \ref{fig:significance}.

\begin{figure*}
\centering
\includegraphics[width=0.8\textwidth]{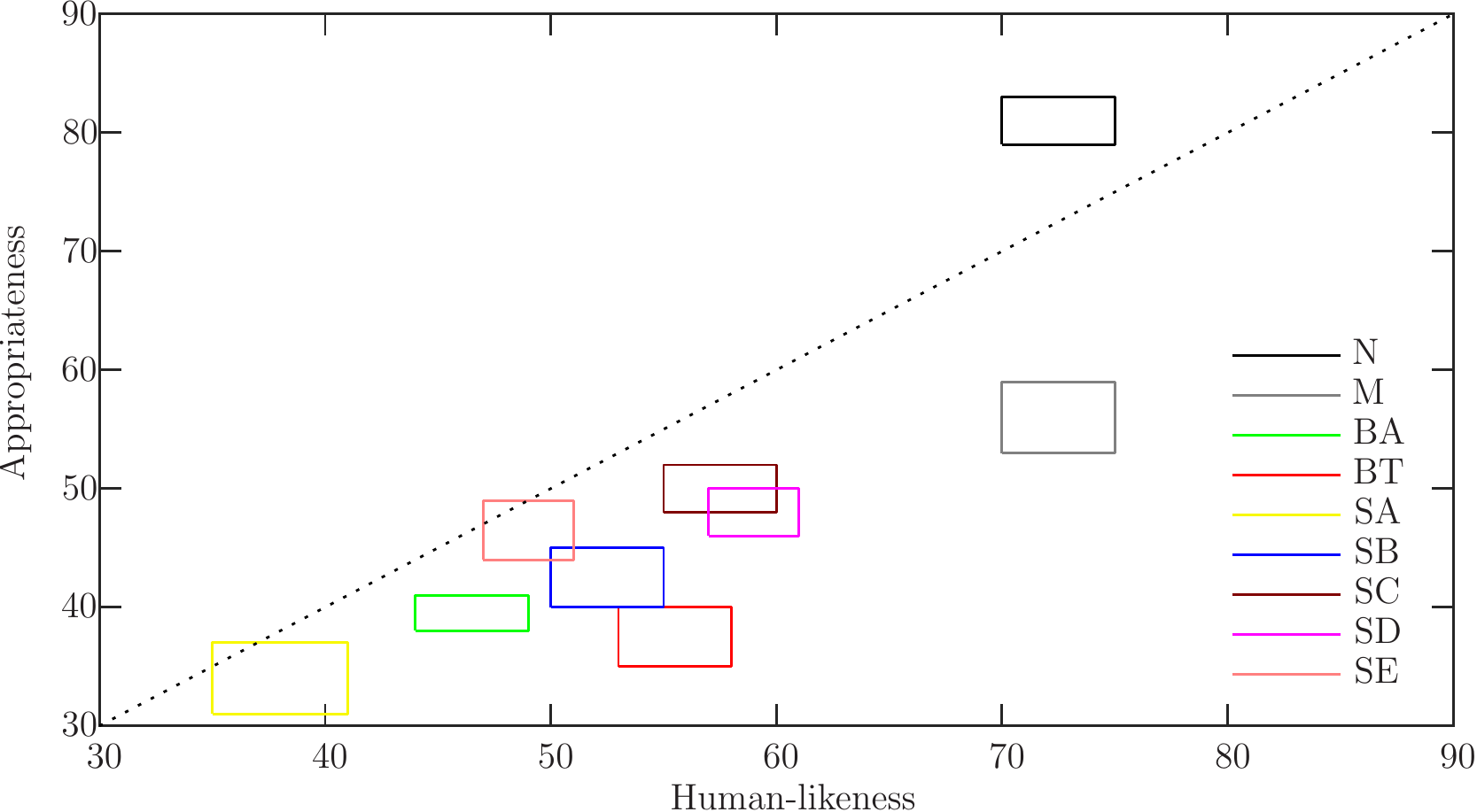}
\caption{Confidence regions for the true median rating across both studies. The dotted black line is the identity, $x=y$. While the human-likeness ($x$-coordinate) of M was not evaluated directly, it is expected to be very close to N since it uses the same motion clips, and the horizontal extent of the confidence region for M was therefore copied from N.}
\label{fig:joint}
\end{figure*}

Finally, we present two diagrams that bring the results of the two studies together.
Figure\ \ref{fig:ordering}, in particular, visualises the relative (partial) ordering between different conditions implied by the results of the two studies in Figure\ \ref{fig:significance}.
Although there are similarities, the two orderings are meaningfully different.
This, together with the results in \cite{jonell2021hemvip},
reinforces a conclusion that the two studies managed to disentangle aspects of perceived motion quality (human-likeness) from the perceived link between gesture and speech (appropriateness).
Figure\ \ref{fig:joint}, meanwhile, visualises confidence regions for the median rating as boxes whose horizontal and vertical extents are given by the corresponding confidence intervals in Table \ref{tab:results}.
Once again, different systems are found to be good at different things.
The numerical gap between natural and synthetic gesture motion is seen to be more pronounced in the case of appropriateness than for human-likeness.



\subsection{Results of objective evaluation}
\label{ssec:objectiveresults}

Results of the objective evaluations from Section\ \ref{subs:obj_metrics} are given in Table~\ref{tab:obj_evaluations}. The first column contains the average jerk across all the joints. We report mean and standard deviation for the full 20 min of test motion. The second and third columns contain the Hellinger distance between speed histograms for the left and right wrists.

Different systems performed best (coming closest to the natural motion N) in different objective measures. For example, systems SA and SB where the closest to the ground truth in terms of the jerk value, but SE and SD were among the closest to the ground truth as measured by Hellinger distance between speed histograms.

We also found that objective metrics deviate from the subjective results. While SA showed the most similar jerk to natural motion, it was less preferred in the subjective evaluation. Similarly, SE showed the Hellinger distances most similar to N, but was not close to being the most preferred synthetic system in the subjective evaluation.  
Considering this disparity, we stress that objective evaluation of gesture motion is a complementary measure, and that subjective evaluation is much more important.

\begin{table}
    \centering
    \caption{Summary statistics of user-study ratings for all conditions in the two studies, with 0.01-level confidence intervals. The human-likeness of M was not evaluated explicitly, since it uses the same motion clips as N.}
    \label{tab:results}
        \begin{tabular}{@{}l|cc|cc@{}}
        \toprule
        & \multicolumn{2}{c|}{Human-likeness} & \multicolumn{2}{c}{Appropriateness}\\
        ID & Median & Mean & Median & Mean\\
        \midrule
        N & $72\in{}[70,\,75]$ & $67.6\pm{}1.8$ & $81\in{}[79,\,83]$ & $73.8\pm{}1.8$\\
        M & \tqdb{} & \tqdb{} & $56\in{}[53,\,59]$ & $53.3\pm{}2.0$\\
        BA & $46\in{}[44,\,49]$ & $46.2\pm{}1.7$ & $40\in{}[38,\,41]$ & $40.4\pm{}1.8$\\
        BT & $55\in{}[53,\,58]$ & $54.6\pm{}1.8$ & $38\in{}[35,\,40]$ & $38.5\pm{}1.9$\\
        SA & $38\in{}[35,\,41]$ & $40.1\pm{}1.9$ & $35\in{}[31,\,37]$ & $36.4\pm{}1.9$\\
        SB & $52\in{}[50,\,55]$ & $52.8\pm{}1.9$ & $43\in{}[40,\,45]$ & $43.3\pm{}2.0$\\
        SC & $57\in{}[55,\,60]$ & $55.8\pm{}1.9$ & $50\in{}[48,\,52]$ & $50.6\pm{}1.9$\\
        SD & $60\in{}[57,\,61]$ & $58.8\pm{}1.7$ & $49\in{}[46,\,50]$ & $48.1\pm{}1.9$\\
        SE & $49\in{}[47,\,51]$ & $49.6\pm{}1.8$ & $47\in{}[44,\,49]$ & $45.9\pm{}1.8$\\
        \bottomrule
        \end{tabular}
\end{table}

\subsection{Discussion of the challenge results}
\label{ssec:discussion_of_results}
It is obvious that gesture generation is a difficult problem which is far from being solved, seeing that no system came remotely close to the natural motion N.
However, the fact that many submissions scored significantly better than the previously published baselines suggests that progress is being made.
The numerical gap between natural motion and that synthesised by machine-learning models is greater in terms of appropriateness than human-likeness. This (along with the fact that no artificial system surpassed the speech-independent condition M) could indicate that appropriateness is a harder problem to solve. As one part of this, the available data may not be sufficiently rich to allow learning to generate appropriate gestures, especially semantically-meaningful gesticulation.

\begin{table}
   \caption{Results from the objective evaluations. The Hellinger distance between natural and synthetic speed profiles was computed for the two wrist joints, since hand motion is of central importance for co-speech gestures.}
    \label{tab:obj_evaluations}
    \centering
        \begin{tabular}{@{}l|ccc@{}}
        \toprule
         &  & \multicolumn{2}{c@{}}{Hellinger distance} \\
        ID & Jerk & Left & Right \\
        \midrule
        N & 151.52 $\pm$ 35.57 & 0\hphantom{.000} & 0\hphantom{.000} \\
         &  &  &  \\
        BA & \hphantom{0}65.59 $\pm$ \hphantom{0}4.42 & 0.084 & 0.090\\
        BT & \hphantom{0}45.84 $\pm$ \hphantom{0}2.14 & 0.130 & 0.096 \\
        SA & 132.37 $\pm$ 27.64 & 0.064 & 0.059 \\
        SB & 189.39 $\pm$ \hphantom{0}4.66 & 0.126 & 0.114\\
        SC & \hphantom{0}84.44 $\pm$ \hphantom{0}8.48 & 0.083 & 0.088 \\
        SD & \hphantom{0}72.06 $\pm$ \hphantom{0}7.91 & 0.073 & 0.062 \\
        SE & \hphantom{0}97.85 $\pm$ \hphantom{0}9.34 & 0.049 & 0.049 \\
        \bottomrule
        \end{tabular}
\end{table}

Previous studies suggest that motion quality (human-likeness) may influence gesture appropriateness ratings in subjective evaluations \cite{yoon2019robots,kucherenko2020gesticulator}.
Our experiments only partly managed to separate these two aspects of gesture perception. On the one hand, we can observe in Figure\ \ref{fig:ordering} that different systems were good at different things: some scored better than other on human-likeness, but worse on appropriateness. 
The human-likeness ratings, which did not include any speech information in the video stimuli, also have little potential to include any aspects of appropriateness. 
On the other hand, no machine-learning system was rated above mismatched motion M in terms of appropriateness, which contrasts against previous evaluations on other data such as \cite{yoon2019robots}.
This could be an effect of that data containing more pauses in speech and gesticulation, thus making a mismatch more apparent.
Moreover, the high appropriateness rating reached by one of the audio-only systems may indicate that our evaluation did not capture semantic appropriateness well.

\section{Discussion and implications of the challenge}
\label{sec:discussion}

In this section we discuss challenge implications: what the challenge brings to the scientific community, the limitations of the challenge, and lessons learned from conducting it.

\subsection{Implications of the challenge}

We have taken the first step in jointly benchmarking different gesture generation systems on a common dataset and virtual avatar.
The below points summarise some of the added value we see for the gesture-generation field:

\begin{enumerate}
    \item We have defined the first benchmark for evaluating gesture-generation models, consisting of a dataset of speech audio, aligned text transcriptions, and 3D motion, as well as train-test splits and an evaluation procedure. Future research can make use of these components to compare new models with previous ones in a consistent way.
    \item All the motion clips generated by the systems evaluated in the challenge are publicly available, together with the rendering pipeline used.\footnote{See \href{https://zenodo.org/record/4080919}{zenodo.org/record/4080919} and \href{https://github.com/jonepatr/genea_visualizer}{github.com/jonepatr/genea\_visualizer} for the motion stimuli and the visualiser, respectively.} This enables easy comparisons with these systems in the future, since their motions can be used directly, without the need to reproduce the systems.
    \item All the subjective and objective scores for the challenge submissions and analysis scripts we used are also available online.\footnote{See \href{https://zenodo.org/record/4088250}{zenodo.org/record/4088250}.} This material could be used, e.g., to investigate human perception and to analyse the correlation between subjective perception and different objective measures (not only those in Section \ref{subs:obj_metrics}), to aid progress toward reliable and useful objective metrics for the field.
\end{enumerate}

\subsection{Limitations}
Our crowdsourced evaluation had a few limitations:
First, in measuring appropriateness of gestures (i.e., the link between gestures and speech), semantic and rhythmic appropriateness were considered together, and there is no way to determine which aspect of appropriateness the participants rated.
In addition, our appropriateness ratings were likely been affected by motion quality to some extent, as discussed in Section\ \ref{ssec:discussion_of_results}, despite the fact that participants were instructed participants to disregard motion quality.

Second, the dataset used in the challenge was limited to a single English speaker in a monologue scenario. 
The role of gesticulation may be expected to differ between different persons and languages as well as the speaking environment (e.g., dyadic conversation versus monologue), which this challenge did not explore. We believe the models and the challenge can be extended to other languages if proper datasets are available, as audio processing is essentially language agnostic and pretrained word vectors are available for a multitude of languages \cite{grave2018learning}.

A third limitation is that we considered only upper-body gestures, even though whole-body gestures (including posture, stepping motion and stance, facial expression, and hand motion) also are important in social interactions. Three teams stated that the most desirable extension of the challenge would be to include whole-body and/or facial gestures. Some evaluation participants also found the absence of facial and finger motion to be a limitation of the challenge.

\subsection{Lessons learned from the challenge}
Conducting the gesture generation challenge has highlighted several take-away messages and lessons learned:
\begin{itemize}
    \item Being human-like does not mean being appropriate for gestures of a virtual avatar. The challenge evaluation found some systems performed better than others in terms of human-likeness but worse in terms of appropriateness, highlighting that one does not imply the other. Any evaluation or comparison of synthetic gestures should keep this distinction in mind.
    \item Providing carefully pre-processed data and good infrastructure (code for feature extraction, motion visualisation, baseline systems, etc.) enables challenge participants to focus on developing their system, instead of solving unrelated issues.
    \item A MUSHRA-like evaluation scheme can successfully benchmark numerous gesture-generation models in parallel.
    \item There is a need for future challenges, since there remains a big gap between natural and synthesised motion and variation across speakers, languages, and scenarios has yet to be explored in a challenge format.
\end{itemize}
We additionally think the following points are worth considering for anyone running a similar challenge in the future:
\begin{itemize}
    \item Include some of the best systems used in the current challenge to provide continuity and assess whether the field keeps moving forward. This is facilitated by the fact that the baselines and several challenge entries have made their code publicly available.
    \item Evaluate gesture appropriateness in a more granular and precise way, for example having separate questions and studies for semantic and rhythmic appropriateness, and by also evaluating contrasts between matched and mismatched motion from all challenge entries. Since the link between speech and motion is important yet difficult to evaluate, challenges and their data may be used to explore how to better measure gesture appropriateness.
    \item Use a different speech-gesture dataset. As previously discussed, the dataset used in this challenge has limitations, e.g., it has already been used extensively and contains just a single actor speaking in isolation, while gesture generation systems usually are intended to be used in an interaction. More data may be necessary to better learn semantically meaningful gestures.
\end{itemize}

\section{Conclusions}
We have hosted the GENEA Challenge 2020 to assess the state of the art in data-driven co-speech gesture generation.
The central design goal of the challenge was to enable direct comparison between many different gesture-generation methods while controlling for factors of variation external to the model, namely data, embodiment, and evaluation methodology.
Our results suggest that the field is advancing measurably, since most submissions performed significantly better than the baselines published the year before.
Different systems were also found to be good at different things on the two scales (human-likeness and appropriateness) that we assessed.
However, a substantial gap remains between synthetic and natural gesture motion, indicating that gesture generation is far from a solved problem.

We believe that the standardised challenge training and test sets (of time-aligned audio, text, and gestures), the visualisation code, and the associated library of rated motion clips from the challenge will be useful for future benchmarking and research in gesture generation.
Furthermore, we think challenges like the one described here are poised to play an important role in identifying key factors for convincing gesture generation in practice, and in driving and validating future progress toward the goal of endowing embodied agents with natural gesture motion.

\begin{acks}
The GENEA Challenge 2020 used the Trinity Speech-Gesture Dataset collected by Ylva Ferstl and Rachel McDonnell. The challenge dataset was further processed by Taras Kucherenko, Simon Alexanderson, Jonatan Lindgren, and Jonas Beskow at KTH Royal Institute of Technology and by Pieter Wolfert at Ghent University.

The authors wish to thank Simon King for sharing his insights and experiences from running the Blizzard Challenge in speech synthesis. We are also grateful to Ulysses Bernardet for input and to Andr{\'e} Tiago Abelho Pereira, Bram Willemsen, Dmytro Kalpakchi, Jonas Beskow, Kevin El Haddad, and Ulme Wennberg for feedback on the paper preprint.

This research was partially supported by Swedish Foundation for Strategic Research contract no.\ RIT15-0107 (EACare), by IITP grant no.\ 2017-0-00162 (Development of Human-care Robot Technology for Aging Society) funded by the Korean government (MSIT), the Flemish Research Foundation grant no.\ 1S95020N, and by the Wallenberg AI, Autonomous Systems and Software Program (WASP) funded by the Knut and Alice Wallenberg Foundation.
\end{acks}

\balance
\bibliographystyle{ACM-Reference-Format}
\bibliography{refs}


\end{document}

%% file: conditions_table.tex
\begin{table*}
\small
\centering
\caption{Conditions participating in the evaluation. Teams are sorted alphabetically by name. The anonymised IDs of submitted entries begin with the letter `S' followed by a second, randomly-assigned letter in the range A through E, but which letter is associated which each team is not revealed in order to preserve anonymity. $\dagger$ indicates a use of word vectors pretrained on external data.}
\begin{tabular}{@{}lll|cc|cc|c@{}}
\toprule 
 &  &  & \multicolumn{2}{c|}{Inputs used} & \multicolumn{2}{c|}{Representation or features} & Stochastic \\
\cmidrule{4-7}
Name or description & Origin & ID & Aud. & Text & Input speech & Motion & output? \\
\midrule
Natural motion & - & N & \cmark & \cmark & -- & -- & \cmark\\
Mismatched motion & - & M & \xmark & \xmark & -- & -- & \cmark\\
\midrule
Audio-only baseline & \citeauthor{kucherenko2019analyzing} \cite{kucherenko2019analyzing} & BA & \cmark & \xmark & MFCC & Exp.\ map & \xmark\\
Text-only baseline & \citeauthor{yoon2019robots} \cite{yoon2019robots} & BT & \xmark & \cmark & FastText${}^\dagger$ & Rot.\ matrix & \xmark\\
\midrule
AlltheSmooth \cite{jinhong_lu_2020_4088376} & CSTR lab, UEDIN, Scotland & S... & \cmark & \xmark & MFCCs & Joint pos.& \xmark\\
Edinburgh CVGU \cite{kunkun_pang_2020_4090879} & CVGU lab, UEDIN, Scotland & S... & \cmark & \cmark & BERT${}^\dagger$ \& mel-spectr. & Rot.\ matrix & \cmark\\
FineMotion \cite{vladislav_korzun_2020_4088609} & ABBYY lab, MIPT, Russia & S...& \cmark & \cmark & GloVe${}^\dagger$ \& mel-spectr. & Exp.\ map & \xmark\\
Nectec \cite{ausdang_thangthai_2020_4088629} & HCCR unit, NECTEC,  & S...& \cmark & \cmark & Phoneme, Spacy word  & Exp.\ map & \xmark\\
& Thailand &&&& vecs.${}^\dagger$, MFCCs, \& prosody &\\
StyleGestures \cite{simon_alexanderson_2020_4088600} & TMH division, KTH, Sweden & S... & \cmark & \xmark & mel-spectr. & Exp.\ map & \cmark\\
\bottomrule
\end{tabular}
\label{tab:conditions}
\end{table*}